\documentclass[12pt]{article}
\usepackage{amssymb}
\usepackage{arxiv}
\usepackage{amsmath}
\usepackage{placeins}
\usepackage{natbib}
\usepackage{caption}
\usepackage{verbatim}
\usepackage{rotating}
\usepackage{geometry}
\usepackage[utf8]{inputenc}
\usepackage{pgfplots}
\usepgfplotslibrary{groupplots}

\pgfplotsset{compat=newest} 

\begin{document}


\title{Virtual plane-wave imaging via Marchenko redatuming}
\author{
  Giovanni Angelo Meles\\
    Department of Geoscience and Engineering \\
      Delft University of Technology \\
      The Netherlands\\
    \texttt{G.A.Meles@tudelft.nl} \\
   \And
  Kees Wapenaar \\
    Department of Geoscience and Engineering \\
      Delft University of Technology \\
      The Netherlands\\
  \texttt{C.P.A.Wapenaar@tudelft.nl} \\
  \And
  Jan Thorbecke \\
    Department of Geoscience and Engineering \\
      Delft University of Technology \\
      The Netherlands\\
  \texttt{Jan.Thorbecke@tudelft.nl} \\
}
\date{}
\maketitle

\section{Abstract}

Marchenko redatuming is a  novel  scheme used to  retrieve up- and down-going Green's functions in an unknown medium.
Marchenko equations are based on reciprocity theorems and are derived on the assumption of the existence of 
functions exhibiting space-time focusing properties once injected in the subsurface.
In contrast to interferometry but similarly to standard migration methods, Marchenko redatuming only requires an estimate of the direct wave from the 
virtual source (or to the virtual receiver), illumination from only one side of the medium, and no physical sources (or receivers) inside the medium. 
In this contribution we consider a different time-focusing condition within the frame of Marchenko redatuming that leads to the retrieval 
of virtual plane-wave responses. As a result, it allows multiple-free imaging using only a one-dimensional sampling of the targeted model at a
fraction of the computational cost of standard Marchenko schemes. The potential of the new method is demonstrated on 2D synthetic models.

\section{Introduction}

Marchenko redatuming estimates Green's functions between the earth's surface and arbitrary locations in the subsurface. 
Differently from seismic interferometry, in Marchenko redatuming no real sources, nor receivers, are required at the chosen subsurface 
locations (\cite{broggini2012focusing} ; \cite{Wapenaar2014}). These Green's functions are evaluated using reciprocity
theorems involving so called 'focusing functions', i.e. wavefields which achieve space-time focusing in
the subsurface.

In principle, redatumed Green's functions can be used to provide
multiple-free images directly (\cite{behura2014autofocus, broggini2014data}). 
However, this approach requires as many virtual sources as there are
image points in the subsurface. Marchenko redatuming
also allows one to perform redatuming of the reflection response from the surface to
a finite number of depth levels and to apply standard imaging in
between those datum levels (\cite{Wapenaar2014}; \cite{ravasi2016target}).
In that case, however, the redatumed reflection responses still include
internal multiples reverberating below the redatuming level, which
again may diminish the quality of resulting images if the distance between the redatuming levels is too large.

Other applications of the Marchenko method include demultiple schemes (\cite{Meles2015,Meles2016,da2017elastic,van2016adaptive}),  microseismic source 
localization (\cite{behura2013imaging, van2017marchenko}), inversion (\cite{Neut2018}) and homogeneous Green's functions retrieval (\cite{urruticoechea2017elastodynamic, Wapenaar2018}).

Despite its requirements on the quality of the reflection response (e.g., knowledge and accurate deconvolution of the source wavelet,  co-location of sources and receiver 
and knowledge of the absolute scaling factor of the recorded data) 
the Marchenko scheme has already been successfully applied to field data (\cite{ravasi2016target, van2015practical,jia2017subsalt,staring2017adaptive,CostaFilho2017}).
Moreover, recent advances have shown how the requirements above can be considerably relaxed by combining the Marchenko equations with a one-way version of the 
Rayleigh integral representation (\cite{ravasi2017rayleigh}).

In this contribution we show how focusing functions associated with virtual plane-wave-responses can be derived 
by imposing a time-focusing condition in the subsurface which allows the derivation of a new set of Marchenko equations.
The virtual plane-wave-responses can be used to efficiently image the subsurface involving only a fraction of virtual-responses 
as compared to standard Marchenko methods. The proposed method thus stands as an ideal bridge between areal-source methods for 
primaries (\cite{rietveld1992}) and the extended virtual-source Marchenko method addressed by \cite{broggini2012ibfocusing}.

Potential and limitations of the new strategy are illustrated by means of numerical examples.

 \section{Method and Theory}

 In this section we briefly introduce reciprocity theorems and use them to derive the coupled Marchenko equations. 
 To simplify our derivations, we will make use of both time and frequency domain expressions. Following standard formalism, we will indicate wavefields
 in the time and frequency domain as $p(\textbf{x},z,t)$ and $\hat{p}(\textbf{x},z,\omega)$, respectively.
 
 Reciprocity theorems for one-way flux-normalized wavefields relate up- and down-going wavefield components of two states A and B evaluated at two depths.
 Convolution and cross-correlation reciprocity theorems can be expressed in the frequency domain as follows (\cite{wapenaar1996reciprocity}):
 
\begin{equation}
 \int_{\Lambda_{a}} d^2 \textbf{x} \{ \hat{p}_A^+\hat{p}_B^- - \hat{p}_A^-\hat{p}_B^+  \} =  \int_{\Lambda_{f}} d^2 \textbf{x} \{ \hat{p}_A^+\hat{p}_B^- - \hat{p}_A^-\hat{p}_B^+  \},
 \label{Convolution}
\end{equation}

\begin{equation}
 \int_{\Lambda_{a}} d^2 \textbf{x} \{ \hat{p}_A^+\hat{p}_B^{+*} - \hat{p}_A^-\hat{p}_B^{-*} \} =  \int_{\Lambda_{f}} d^2 \textbf{x} \{ \hat{p}_A^+\hat{p}_B^{+*} - \hat{p}_A^-\hat{p}_B^{-*}  \},
\label{Crosscorrelation}
 \end{equation}

where $*$ is complex conjugation, subscripts $A$ and $B$ relate to the corresponding 
states, superscripts $+$ and $-$ indicate down- and up-going constituents, and $\Lambda_{a}$ and $\Lambda_{f}$ stand for two arbitrary depth levels.

Equations (\ref{Convolution}) and (\ref{Crosscorrelation}) assume that the  medium parameters are identical 
for both states in the volume circumscribed by $\Lambda_{a}$ and $\Lambda_{f}$, and that no sources exist between these depth levels.
Moreover, while (\ref{Convolution}) is valid for lossy media, (\ref{Crosscorrelation}) requires the medium to be lossless between the levels
$\Lambda_{a}$ and $\Lambda_{f}$, thus posing a limitation to the methodology presented here (for an extension to account for dissipation see \cite{slob2016green}). Moreover, evanescent waves
are neglected in equation  (\ref{Crosscorrelation}).

We will consider $\Lambda_a$ and $\Lambda_f$ to be the acquisition surface and a redatuming level, respectively.
Moreover, we consider for state A a truncated medium identical to the physical medium above $\Lambda_f$ and reflection-free below this level, 
while for state B we choose the physical medium.

We now discuss and define the properties of the wavefield for state A for two different focusing conditions, which we will refer to as $f$ and $F$. 

\begin{figure} 
  \centering
   \includegraphics[width=0.95\textwidth]{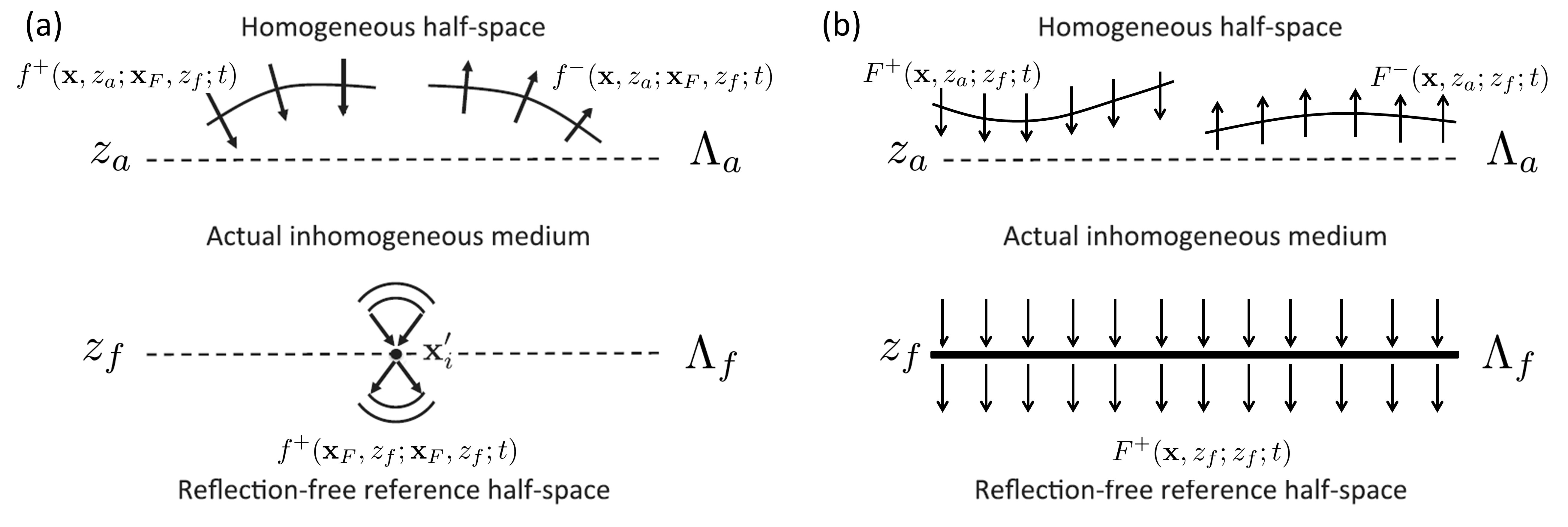} 
   \caption{Down- and up-going components of the focusing functions of the 3D wave equation in a reference configuration. 
   (a) Standard space-time focusing function $f_1$, leading to point source responses (Green's functions).
(b) Time focusing function $F_1$, leading to areal-source-responses. 
}
  \label{Focusing_Conditions}
  \end{figure}

In standard space-time focusing, it is assumed that the down-going component of the focusing function representing state A, i.e. $f_1^+$, satisfies the following focusing condition along
\mbox{$\Lambda_f:$} 
$f_1^+(\textbf{x},z_f;\textbf{x}_F,z_f;t) = \delta(t)\delta(\textbf{x}-\textbf{x}_F)$, where $\textbf{x}_F,z_f$ are  
the coordinates of a focal point in the subsurface (in the frequency domain this becomes: \mbox{$\forall \omega, \hat{f}_1^+(\textbf{x},z_f;\textbf{x}_F,z_f;\omega) = \delta(\textbf{x}-\textbf{x}_F)$}). 
Moreover, since it is assumed that the medium in state $A$ is truncated 
below $\Lambda_f$, no up-going component $f_1^-$ exists along this lower boundary regardless of its properties along $\Lambda_a$ (Figure \ref{Focusing_Conditions}(a)).

For state B, following the standard approach, we place a point source for a downgoing wavefield at $\textbf{x}_B$ at depth $z_a$ just above the surface, 
so that along $\Lambda_a$ we have $\hat{p}_B^+=\delta(\textbf{x}-\textbf{x}_B)$ and $\hat{p}_B^-=\hat{R}(\textbf{x},z_a;\textbf{x}_B,z_a,\omega)$,
where $\hat{R}$ indicates the reflection response of the physical medium at the surface, 
and on $\Lambda_f$ we have $p_B^{+/-} = g^{+/-}(\textbf{x},z_f;\textbf{x}_B,z_a,\omega)$, where $g^+$ and $g^-$ are the down- and up-going parts of the Green's function.

Substituting these definitions into equations (\ref{Convolution}) and (\ref{Crosscorrelation}) we get:

\begin{equation}
\begin{aligned}
 \hat{f}_1^-(\textbf{x}_B,z_a;\textbf{x}_F,z_f;\omega) +  \hat{g}^-(\textbf{x}_F,z_f;\textbf{x}_B,z_a;\omega)   =  \\
 \int_{{\Lambda_a}} d^2 \textbf{x} \hat{R} (\textbf{x},z_a;\textbf{x}_B,z_a;\omega) \hat{f}_1^+(\textbf{x},z_a;\textbf{x}_F,z_f;\omega),    \\
 \hat{f}_1^+(\textbf{x}_B,z_a;\textbf{x}_F,z_f;\omega) -  \hat{g}^{+*}(\textbf{x}_F,z_f;\textbf{x}_B,z_a;\omega)   =  \\
  \int_{{\Lambda_a}} d^2 \textbf{x} \hat{R^*} (\textbf{x},z_a;\textbf{x}_B,z_a;\omega) \hat{f}_1^-(\textbf{x},z_a;\textbf{x}_F,z_f;\omega),  \\
\end{aligned}
 \end{equation}

or, using the compact, time-domain formalism introduced in \cite{vanderNeutetal2015}:

\begin{equation}
\begin{aligned}
 \textbf{f}_\textbf{1}^\textbf{-} + \textbf{g}^\textbf{-} = \textbf{Rf}_\textbf{1}^\textbf{+},  \\
 \textbf{f}_\textbf{1}^\textbf{+} - \textbf{g}^{\textbf{+}\star} = \textbf{R}^{\star}\textbf{f}_\textbf{1}^\textbf{-},\\
 \label{coupledMarchenko}
\end{aligned}
\end{equation}

where the superscript $\star$ indicates time-reversal. 

We now analyze this standard problem in more detail, and show how the algorithm that provides its solution can be easily extended to problems involving different focusing conditions. 
The underdetermined system in equation (\ref{coupledMarchenko}), which represents the basis for standard Marchenko redatuming, can be additionally 
simplified invoking a separation operator $\Theta_f$ to annihilate the Green's functions terms:

\begin{equation}
\begin{aligned}
\Theta_f \textbf{f}_\textbf{1}^\textbf{-} = \Theta_f \textbf{R} \textbf{f}_\textbf{1}^\textbf{+},  \\
 \Theta_f \textbf{f}_\textbf{1}^\textbf{+} = \Theta_f \textbf{R}^{\star}\textbf{f}_\textbf{1}^\textbf{-}.\\
 \label{SimplifiedMarchenko}
\end{aligned}
\end{equation}
  
This leads, after decomposing the focusing function into a direct and a coda component
(i.e., setting $\textbf{f}_\textbf{1}^\textbf{+} = \textbf{f}_{\textbf{1d}}^\textbf{+} + \textbf{f}_{\textbf{1m}}^\textbf{+}$, 
and using $\Theta_f \textbf{f}_\textbf{1}^\textbf{-}=\textbf{f}_\textbf{1}^\textbf{-}$ and $\Theta_{f} \textbf{f}_\textbf{1}^\textbf{+}=\textbf{f}_{\textbf{1m}}^\textbf{+}$), to the invertible 
linear problem
 
\begin{equation}
 [ \textbf{I} - \Theta_f \textbf{R}^{\star}\Theta_f \textbf{R} ] \textbf{f}_{\textbf{1m}}^\textbf{+} = \Theta_f \textbf{R}^{\star}\Theta_f \textbf{R} \textbf{f}_{\textbf{1d}}^+,
 \label{Linear_Equation}
\end{equation}

which, under standard convergence conditions (\cite{Fokkema2013}), is solved by:
\begin{equation}
 \textbf{f}_\textbf{1}^{\textbf{+}} = \sum_{k=0}^\infty (\Theta_f \textbf{R}^{\star} \Theta_f \textbf{R})^k \textbf{f}_{\textbf{1d}}^+.
 \label{MarchenkoSolution}
\end{equation}

Once the focusing functions are found, they are inserted in equation \ref{coupledMarchenko}, yielding the point source Green's functions $\textbf{g}^{\textbf{-}}$ and $\textbf{g}^{\textbf{+}}$.

More details about the derivation of this series solution can be found in \cite{vanderNeutetal2015}, while in \cite{Dukalski2017} other algorithms to solve equation (\ref{Linear_Equation}) are analysed.

We now consider a different focusing condition, which we will refer to as 'time-focusing condition' and show how its imposition results in the same Marchenko equations discussed above. 
For the time-focusing approach we refer to the focusing wavefield in state $A$ as $F_1$.  We assume $F_1$ to be defined in a medium truncated 
below $\Lambda_f$, and therefore also in this case no up-going component $F_1^-$ exists along this lower boundary regardless of its properties along $\Lambda_a$.
However, differently from the standard space-time focusing approach, we define $F_1^+$ as satisfying the following time-focusing condition
along \mbox{$\Lambda_f: \forall \textbf{x} \in \Lambda_f, F_1^+(\textbf{x},z_f;z_f,t) = \delta(t)$}, where $z_f$ is the depth of the horizontal focal plane
in the subsurface.

Note that in the frequency domain this becomes: \mbox{$ \forall \textbf{x} \in \Lambda_f, \forall \omega, \hat{F}_1^+(\textbf{x},z_f;z_f,\omega) = 1$}.
Note also that the time-focusing condition can be interpreted as a spatial integral along $\Lambda_f$ of space-time focusing conditions, namely:

\begin{equation}
F_1^+(\textbf{x},z;z_f;t) = \int_{\Lambda_f} d^2 \textbf{x}_F f_1^+(\textbf{x},z;\textbf{x}_F,z_f;t).
\end{equation}

It is therefore clear that the focusing function $F_1$ could be obtained by integrating an appropriate set of focusing functions $f_1$, each involving
the solution of a Marchenko equation (see equation \ref{MarchenkoSolution}).
We will show in the following that the solution of a single Marchenko equation can provide the same result.

For state B we consider again a point source for a downgoing wavefield at $\textbf{x}_B$ just above the surface.

Substituting these definitions into equations (\ref{Convolution}) and (\ref{Crosscorrelation}) we get:

\begin{equation}
\begin{aligned}
 \hat{F}_1^-(\textbf{x}_B,z_a;z_f;\omega) +   \hat{G}^-(z_f;\textbf{x}_B,z_a;\omega)   =  \\
 \int_{{\Lambda_a}} d^2 \textbf{x} \hat{R} (\textbf{x},z_a;\textbf{x}_B,z_a;\omega) \hat{F}_1^+(\textbf{x},z_a;z_f;\omega),    \\
 \hat{F}_1^+(\textbf{x}_B,z_a;z_f;\omega) -  \hat{G}^{+*}(z_f;\textbf{x}_B,z_a;\omega)   =  \\
  \int_{{\Lambda_a}} d^2 \textbf{x} \hat{R^*} (\textbf{x},z_a;\textbf{x}_B,z_a;\omega) \hat{F}_1^-(\textbf{x},z_a;z_f;\omega).    \\
   \label{ComplexMarchenko}
\end{aligned}
 \end{equation}
 
 where
 
 \begin{equation}
\begin{aligned}
\hat{G}^-(z_f;\textbf{x}_B,z_a;\omega)  = \int_{{\Lambda_f}} d^2 \textbf{x} \hat{g}^-(\textbf{x}, z_f;\textbf{x}_B,z_a; \omega) , \\
\hat{G}^{+*}(z_f;\textbf{x}_B,z_a;\omega)   = \int_{{\Lambda_f}} d^2 \textbf{x} \hat{g}^{+}(\textbf{x}, z_f;\textbf{x}_B, z_a;\omega), \\
 \label{G_def}
\end{aligned}
\end{equation}
 
or, using again the compact, time-domain formalism:
 
\begin{equation}
\begin{aligned}
 \textbf{F}_\textbf{1}^\textbf{-} + \textbf{G}^\textbf{-} = \textbf{RF}_\textbf{1}^\textbf{+},  \\
 \textbf{F}_\textbf{1}^\textbf{+} - \textbf{G}^{\textbf{+}\star} = \textbf{R}^{\star}\textbf{F}_\textbf{1}^\textbf{-}.\\
 \label{ExcoupledMarchenko}
\end{aligned}
\end{equation} 

Similarly to what considered in equation (\ref{coupledMarchenko}), the set of equations in (\ref{ExcoupledMarchenko}) is also underdetermined.
As discussed above, the key ingredient to solve the system in equation (\ref{coupledMarchenko}) is the existence of
an appropriate separation operator. 

Such an operator does not necessarily exist only for the space-time focusing system (\ref{coupledMarchenko}), as already 
preliminarily observed in \cite{broggini2012ibfocusing} for slightly spatially-extended virtual sources.
Here we generalize the observation of \cite{broggini2012ibfocusing}, now considering plane-wave
spatially-extended sources. More precisely, we postulate that when a focusing function $\textbf{F}_\textbf{1}^\textbf{+} $ 
satisfies the time-focusing property discussed above, a separation operator $\Theta_F$ (based on the kinematics of the response 
of $\int_{{\Lambda_f}} d^2 \textbf{x} g^+_d( \textbf{x},z_f, t;\textbf{x}', z_a,0)$, which can be interpreted as a plane-wave source based on reciprocity) 
can be successfully applied to equation (\ref{ExcoupledMarchenko}).

In this scenario, the existence of a separation operator reduces (\ref{ExcoupledMarchenko}) into:

\begin{equation}
\begin{aligned}
 \Theta_F \textbf{F}_\textbf{1}^\textbf{-} = \Theta_F \textbf{RF}_\textbf{1}^\textbf{+},  \\
 \Theta_F \textbf{F}_\textbf{1}^\textbf{+} = \Theta_F \textbf{R}^{\star}\textbf{F}_\textbf{1}^\textbf{-}.\\
 \label{PMarchenko}
\end{aligned}
\end{equation}

Following again the decomposition into a direct and coda component of the down-going focusing function, this leads to the solution for the focusing function:

\begin{equation}
 \textbf{F}_\textbf{1}^{\textbf{+}} = \sum_{k=0}^\infty (\Theta_F \textbf{R}^{\star} \Theta_F \textbf{R})^k \textbf{F}_{\textbf{1d}}^\textbf{+}.
 \label{PMarchenkoSolution}
\end{equation}

Once the focusing functions are found, they are inserted in equation (\ref{ExcoupledMarchenko}), yielding the plane-wave source 
Green's functions $\hat{G}^-(z_f;\textbf{x}_B,z_a;\omega)$ and $\hat{G}^{+*}(z_f;\textbf{x}_B,z_a;\omega)$.
This scheme therefore results in the
retrieval of plane wave \textit{up-} and \textit{down-}going areal-receiver-responses (by invoking reciprocity, 
these responses can be related to the \textit{down-} and \textit{up-}propagating areal-sources-responses 
discussed in \cite{rietveld1992})
rather than standard \textit{up-} and \textit{down}-going Green's functions as in \cite{vanderNeutetal2015}.

Once these plane-wave-responses are available, they could be used within the areal sources framework (\cite{rietveld1992}).

 \section{Numerical Examples}

 \subsection{Focusing performances}
 
  \begin{figure} 
  \centering
   \includegraphics[width=0.95\textwidth]{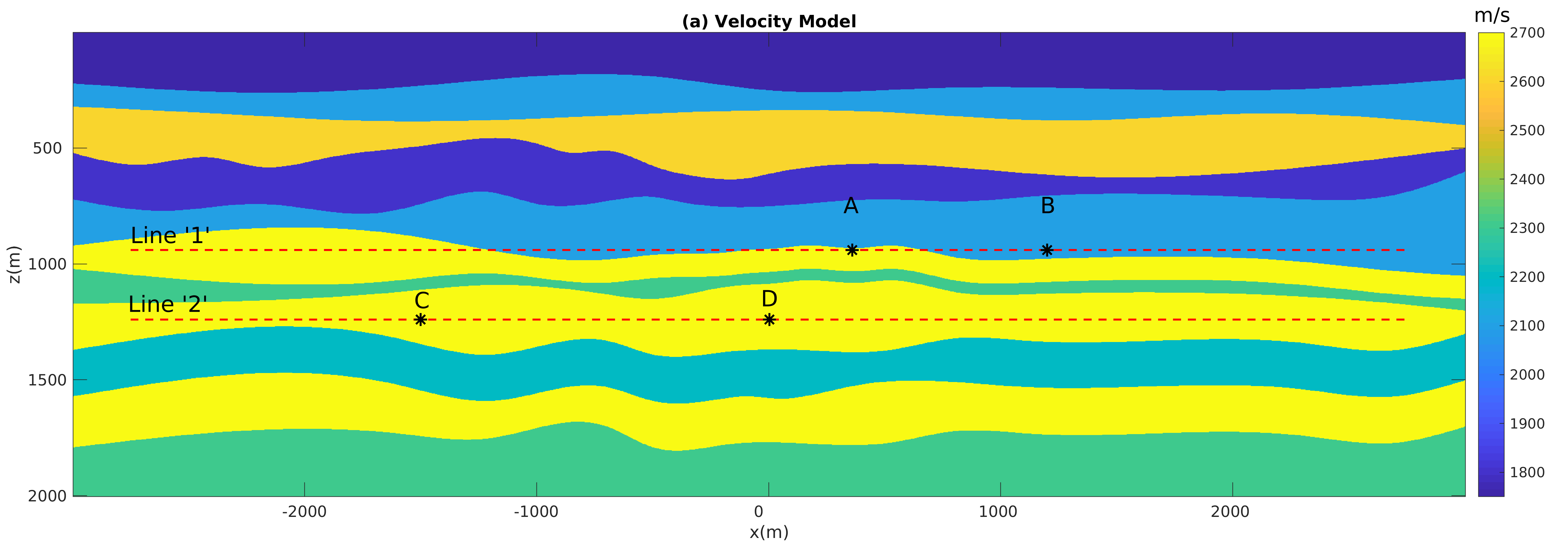}
   \includegraphics[width=0.95\textwidth]{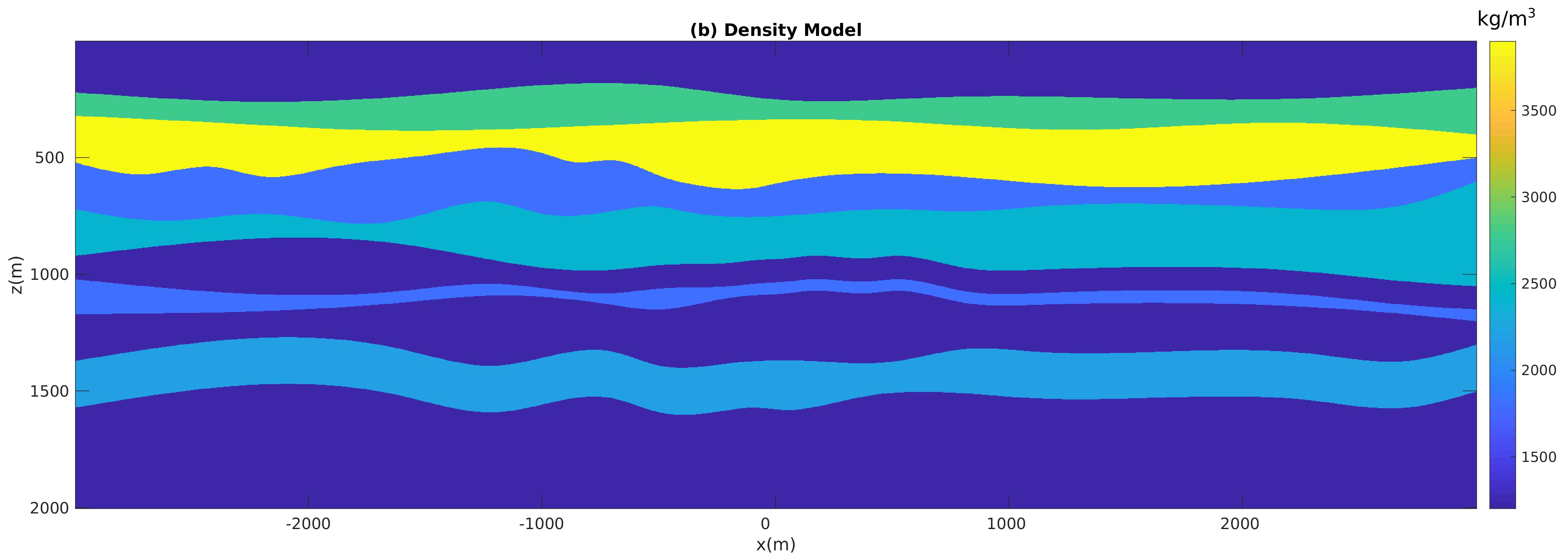}  
   \includegraphics[width=0.95\textwidth]{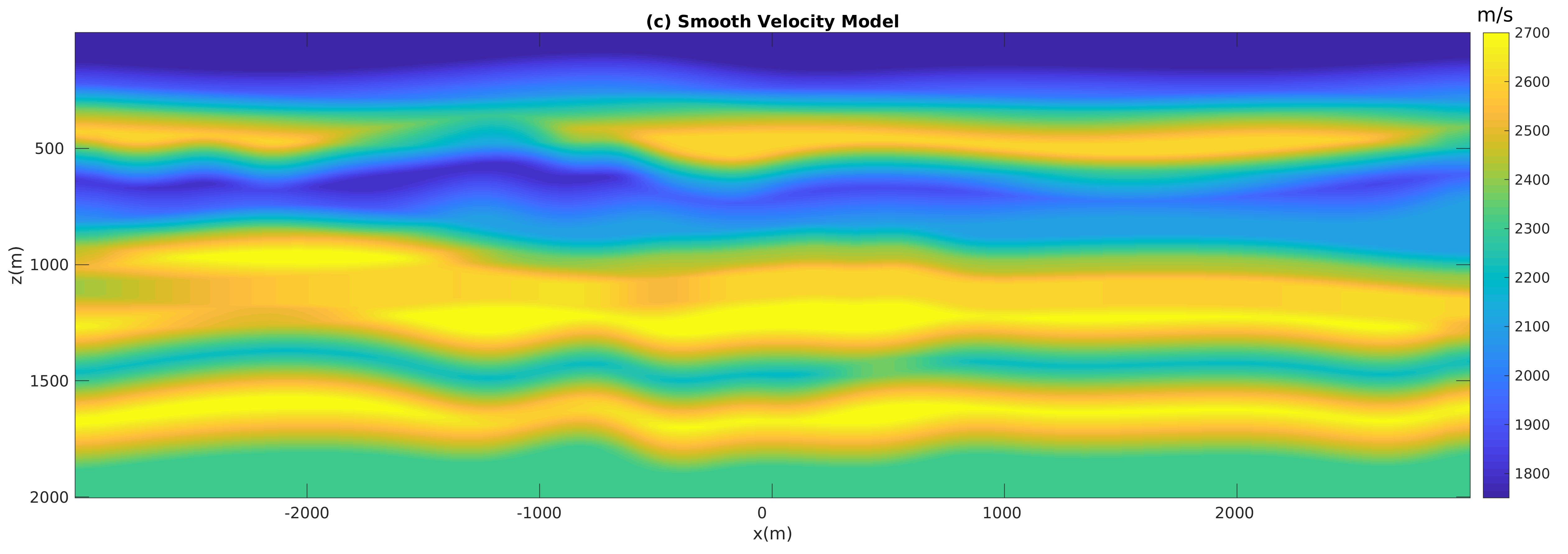}
   \includegraphics[width=0.95\textwidth]{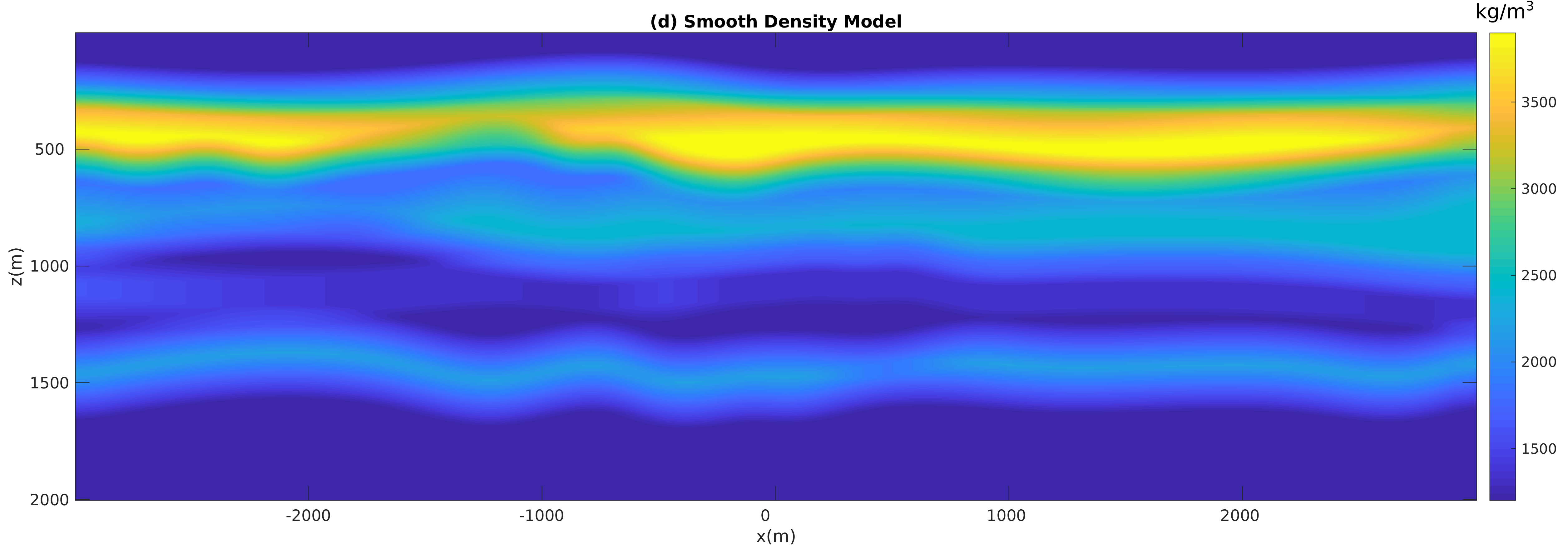}  
\caption{(a) Velocity model used in the first numerical experiment.
Dashed lines and stars represents subsurface planes and points for time and space-time focusing, respectively.(b) Density model used in the numerical experiment. 
(c) and (d) Smooth Velocity and Density models used to provide input for Marchenko redatuming.
}
  \label{true_model}
  \end{figure}
 
We illustrate the potential of the iterative solutions algorithm for areal-sources-responses with 
Finite Difference examples (\cite{thorbecke2017implementation}). We consider the 2D inhomogeneous subsurface model in Figure
\ref{true_model}.

First we assess the focusing performances of the solution of (\ref{MarchenkoSolution}) when a separation operator $\Theta_F$ and an initial
focusing function $\textbf{F}_{\textbf{1d}}^\textbf{+}$ associated with the first arrival of
$\int_{{\Lambda_f}} d^2 \textbf{x} g^+(\textbf{x}, z_f;t,\textbf{x}', z_a;0 )$)  are used. 

\begin{figure} 
  \centering
   \includegraphics[width=0.65\textwidth]{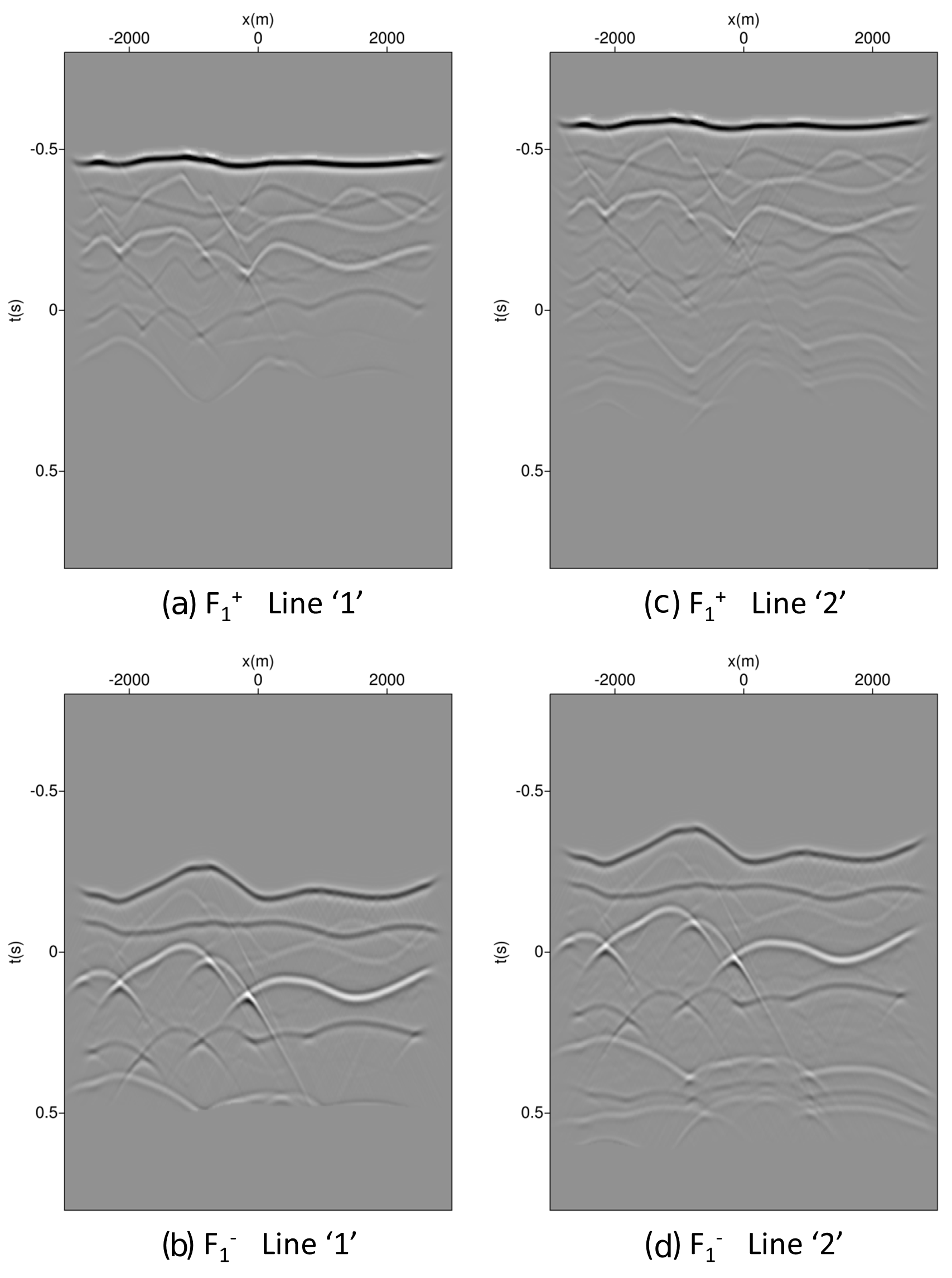} 
   \caption{(a) Down-going component of the Focusing function $F_1$ associated with $F_{1d^+}$ being first arrival of $ \int_{{L_1}} d^2 \textbf{x} g^+(\textbf{x}, z_f;t,\textbf{x}', z_a;0)$.
   (b) Up-going component of the Focusing function $F_1$ associated with $F_{1d^+}$ being the first arrival of $\int_{{L_1}} d^2 \textbf{x} g^+(\textbf{x}, z_f;t,\textbf{x}', z_a;0)$. (c) and (d): as for (a) and
   (b), but for $F_{1d^+} $ associated with the first arrival of $\int_{{L_2}} d^2 \textbf{x} g^+(\textbf{x}, z_f;t,\textbf{x}', z_a;0 )$.}
  \label{Focusing_Examples}
  \end{figure}

We consider two arbitrarily chosen different depth levels (Lines '1' and '2' in Figure \ref{true_model}).
We then solve equation (\ref{PMarchenkoSolution}) 
for initial focusing functions $\textbf{F}_{\textbf{1d}}^\textbf{+}$ related to the depth levels of Lines '1' and '2', respectively, computing these direct components 
using the smooth models in Figure \ref{true_model}(c) and (d). The resulting up- and down-going focusing functions are shown in Figure \ref{Focusing_Examples}.
We then inject the retrieved downgoing focusing functions $\textbf{F}_{\textbf{1}}^\textbf{+}$ (Figures \ref{Focusing_Examples}(a) and (c))
into the corresponding truncated media, and record their response along Lines '1' and '2', respectively. 

\begin{figure} 
  \centering
  \includegraphics[width=0.95\textwidth]{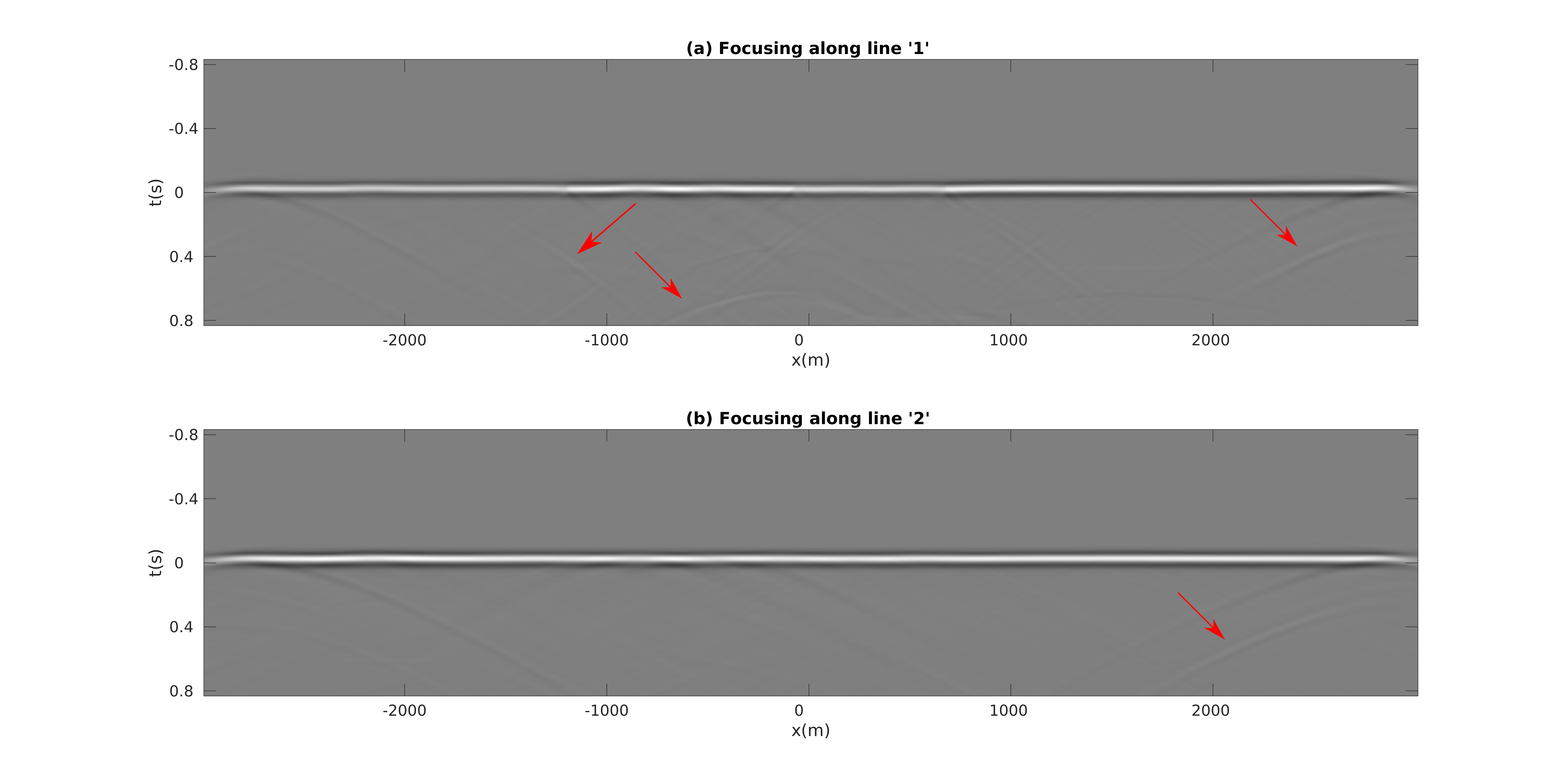}
  \caption{(a) Time-Focusing along Line '1' in Figure \ref{true_model}(a). (b) Time-Focusing along line '2' in Figure \ref{true_model}(a). Red arrows point at small-amplitude artefacts.}
  \label{Plane_Focusing}
  \end{figure}

\begin{figure} 
  \centering
  \includegraphics[width=0.95\textwidth]{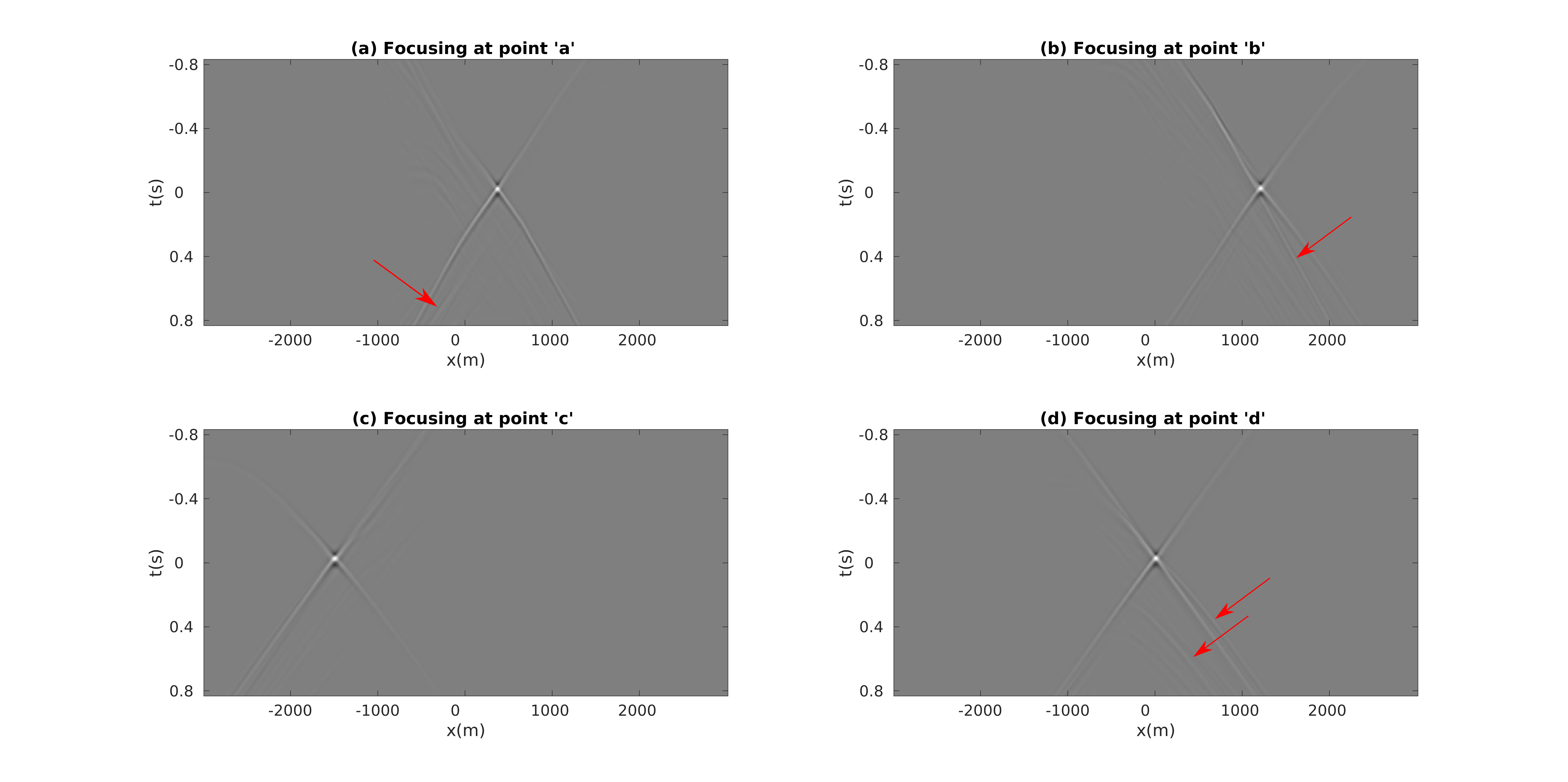}
  \caption{(a) Space-Time Focusing at point 'a' in Figure \ref{true_model}. (b)-(d), as for (a), but for points (b)-(d) in Figure \ref{true_model}. Red arrows point at small-amplitude artefacts.}
  \label{Point_Focusing}
  \end{figure}

Figure \ref{Plane_Focusing} shows that for both cases the focusing is very good, with only small 
amplitude artefacts contaminating the wavefield
along the focal plane (red arrows in Figure \ref{Plane_Focusing}). 
Note that Line '1' crosses an interface, and therefore represents a particularly challenging problem due to the intrinsic limitations of the Marchenko method 
at interfaces, where the validity of the separation operator can be violated (\cite{Vasconcelos2014}).
The overall focusing performances are comparable to those 
of the standard Marchenko method, shown in Figure (\ref{Point_Focusing}) for representative points located close to or 
far-away from interfaces ((a)-(d) in Figure \ref{true_model}), 
where artefacts (partially due to the finite acquisition aperture) are also seen to contaminate the focusing (red arrows in Figure \ref{Point_Focusing}). Note 
that smooth models (see Figure \ref{true_model}(c) and (d)) were used to initiate the focusing process, 
and that perfect foci cannot be expected.\\

Figure \ref{Areal_Solutions} compares directly modelled and retrieved Marchenko  areal-source-responses at the surface for Lines '1' and '2', respectively. As a direct consequence of the excellent
focusing performances demonstrated in Figure \ref{Plane_Focusing}, the match between the modelled and the retrieved areal-responses is also very good,
with mainly tapering-related minor differences in the left- and right-most portions of the gather. 

  \begin{figure} 
  \centering
  \includegraphics[width=0.95\textwidth]{Gathers.png}
  \caption{(a) FD modelled (black lines) and Marchenko (red lines) areal source responses for Line '1'. (b) As for (a), but for Line '2'.}
  \label{Areal_Solutions}
  \end{figure}

\subsection{Imaging results}

As mentioned in the introduction, redatumed Green's functions can be used to provide
multiple-free images directly, by cross-correlation of up- and direct down-going wavefields in the subsurface (\cite{behura2014autofocus}).
However, this approach is expensive, as it requires as many virtual sources as there are
image points in the subsurface (number of required Marchenko solutions: $nx \times nz$ in 2D, or $nx \times ny \times nz$ in 3D, where $nx$, $ny$ and $nz$ stand for the number
of image points along the $x$, $y$ and $z$ axis, respectively).

  \begin{figure} 
  \centering

   \includegraphics[width=0.95\textwidth]{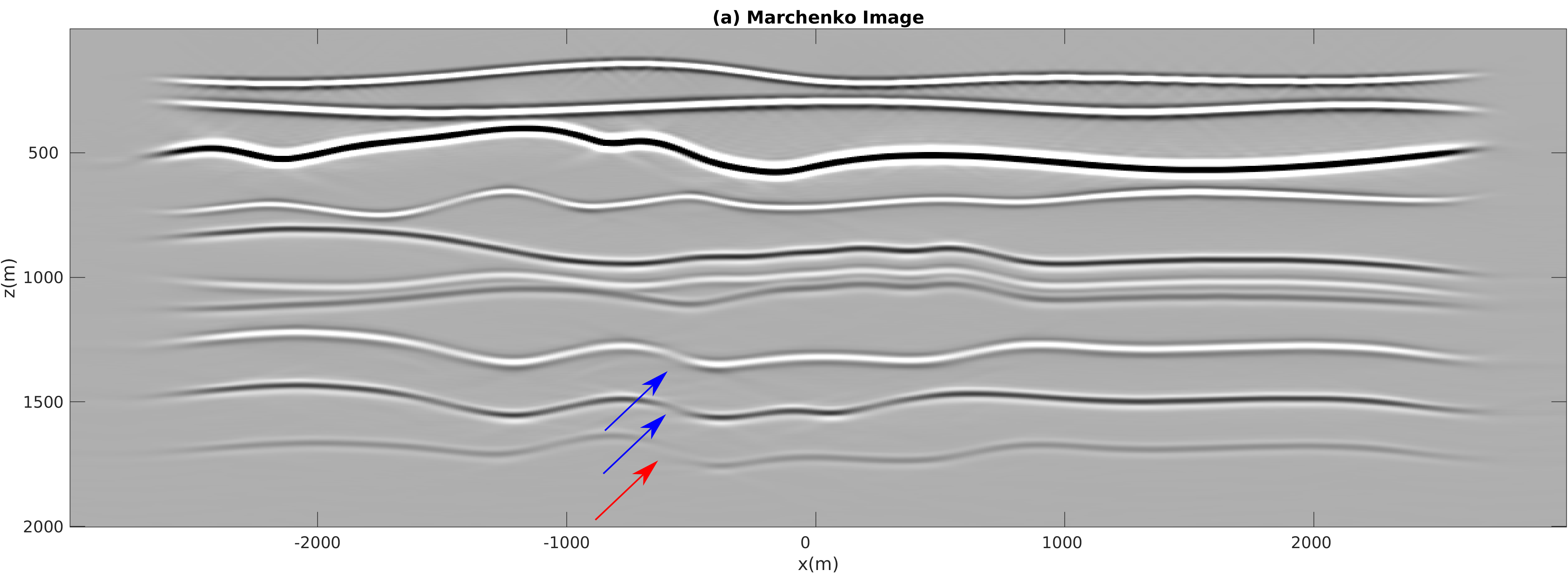} 
   \includegraphics[width=0.95\textwidth]{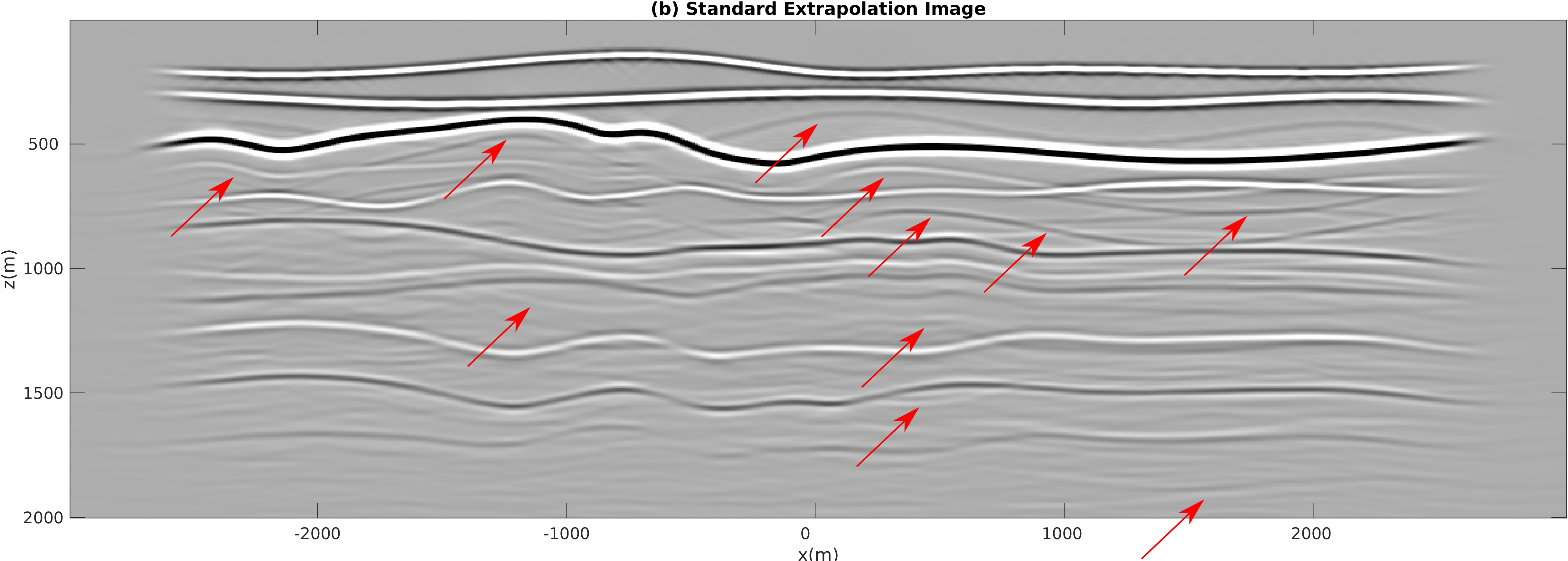} 
\caption{(a) Migration result using the imaging
condition of equation (\ref{migration_imaging}) and Marchenko redatumed virtual-plane wavefields. The red arrow points at a poorly imaged dipping layer, whereas the blue arrows point at similar structures that are properly imaged. 
(b) Migration result using standard one-way extrapolation of virtual-plane wavefields. Red arrows point at multiple-related artefacts.}
  \label{Imaging_1}
  \end{figure}

With Marchenko areal-source-responses, however, we can use a single redatumed solution to image a whole line/plane at once (number of required Marchenko solutions: $nz$ in 2D as well as in 3D).
To achieve this, we use the following redatumed reflectivity and standard migration imaging condition definitions, in the frequency domain:

\begin{equation}
\begin{aligned}
\hat{R}(\textbf{x},z_f,\omega) &= \int_{\Lambda_a} d^2 \textbf{x}' \hat{g_d}^{+*} (\textbf{x}, z_f, \textbf{x}',z_a,\omega) \hat{G}^{-}(z_f;\textbf{x}',z_a,\omega), \\
I(\textbf{x},z_f) &= \int_{\mathbb{R}}d\omega \hat{R}(\textbf{x},z_f,\omega) . \\
\label{migration_imaging}
\end{aligned}
\end{equation}

Note that the imaging condition in (\ref{migration_imaging}) can be seen as an integral along the focal plane of point sources imaging conditions (this integration is implicit in $\hat{G}^{-}$).
We expect this integration to reduce the lateral resolution of the final image due to poorer angle-illumination. In the following we will present a strategy to account for this limitation.

We apply our new imaging condition to the model discussed in the previous section. In this case we sample in depth every 5 meters, and consequently to image the entire domain 
we employ 400 virtual areal sources. Note that imaging the whole model at a 5 meters sampling rate using other Marchenko schemes would involve the computation and migration of up to 400*1200
virtual point-source responses, which would require considerable CPU or RAM resources. An exhaustive analysis about the computational burden of the Marchenko method 
as an imaging tool can be found in \cite{behura2014autofocus}.
The migration associated with the imaging condition in equation (\ref{migration_imaging}) is shown in Figure \ref{Imaging_1}(a). 
Each interface is properly imaged, while no multiple-related artefacts are present. Only a dipping layer in the bottom of the model is relatively poorly 
imaged (red arrow in Figure \ref{Imaging_1}(a)), partially due to its smaller impedance contrast. In any case, others structures with similar geometry are properly imaged 
(blue arrows in Figure \ref{Imaging_1}(a)).
Multiple-related artefacts, on the other hand, contaminate the 
image if we migrate the up-going response associated with the same areal sources obtained through standard one-way wavefield extrapolation (Figure \ref{Imaging_1}(b)).
Note that in the migration step the same smooth models 
depicted in Figure \ref{true_model}(c) and (d) employed for Marchenko redatuming were used.
\\

\begin{figure} 
\centering
\includegraphics[width=0.95\textwidth]{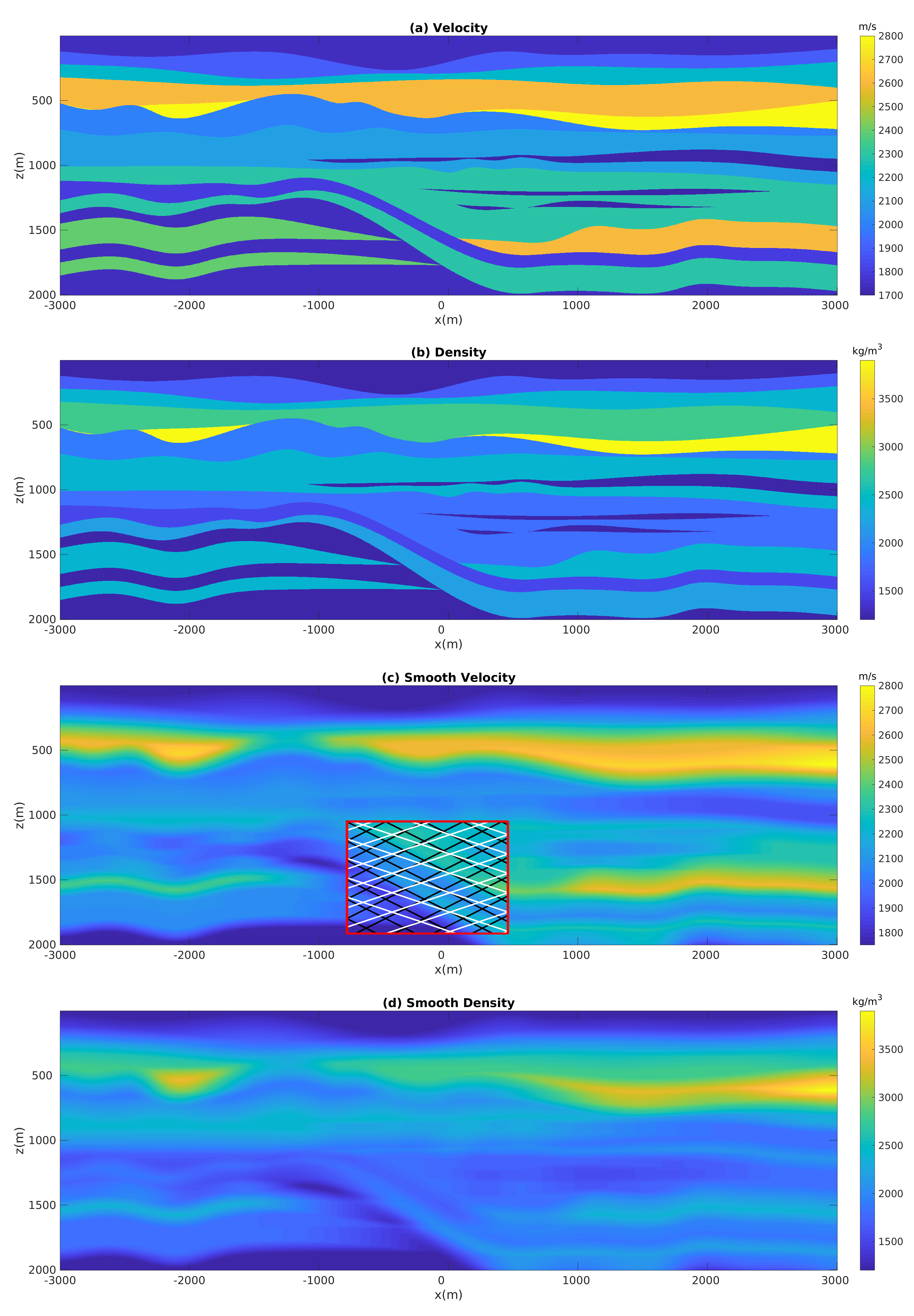}
\caption{(a) Velocity model used in the second numerical experiment.
(b) Density model used in the second numerical experiment. 
(c) and (d) Smooth Velocity and Density models used to provide input for Marchenko redatuming. The red box in (c) indicates a sub-zone of the model where tilted planes, represented by white and black lines, are used
to improve the final imaging result.
}
\label{2D_model}
\end{figure}

We further investigate the potential and limitations of the Marchenko plane wave imaging scheme by considering a more complex subsurface model (Figure \ref{2D_model}(a) and (b)).
Differently from the medium of the first experiment (Figure \ref{true_model}), the model considered here comprises dipping structures and diffractors.
We follow the same imaging strategy discussed for the first numerical experiment, i.e. we initially compute Marchenko areal-source-responses for a set of evenly spaced 
(sampling every 5 meters in depth) horizontal boundaries and we then apply the redatuming and imaging condition discussed in equation (\ref{migration_imaging}).
As for the previous experiment, we employ smooth velocity and density distributions both for the Marchenko and migration steps (Figure \ref{2D_model}(c) and (d)).

\begin{figure} 
\centering
\includegraphics[width=0.95\textwidth]{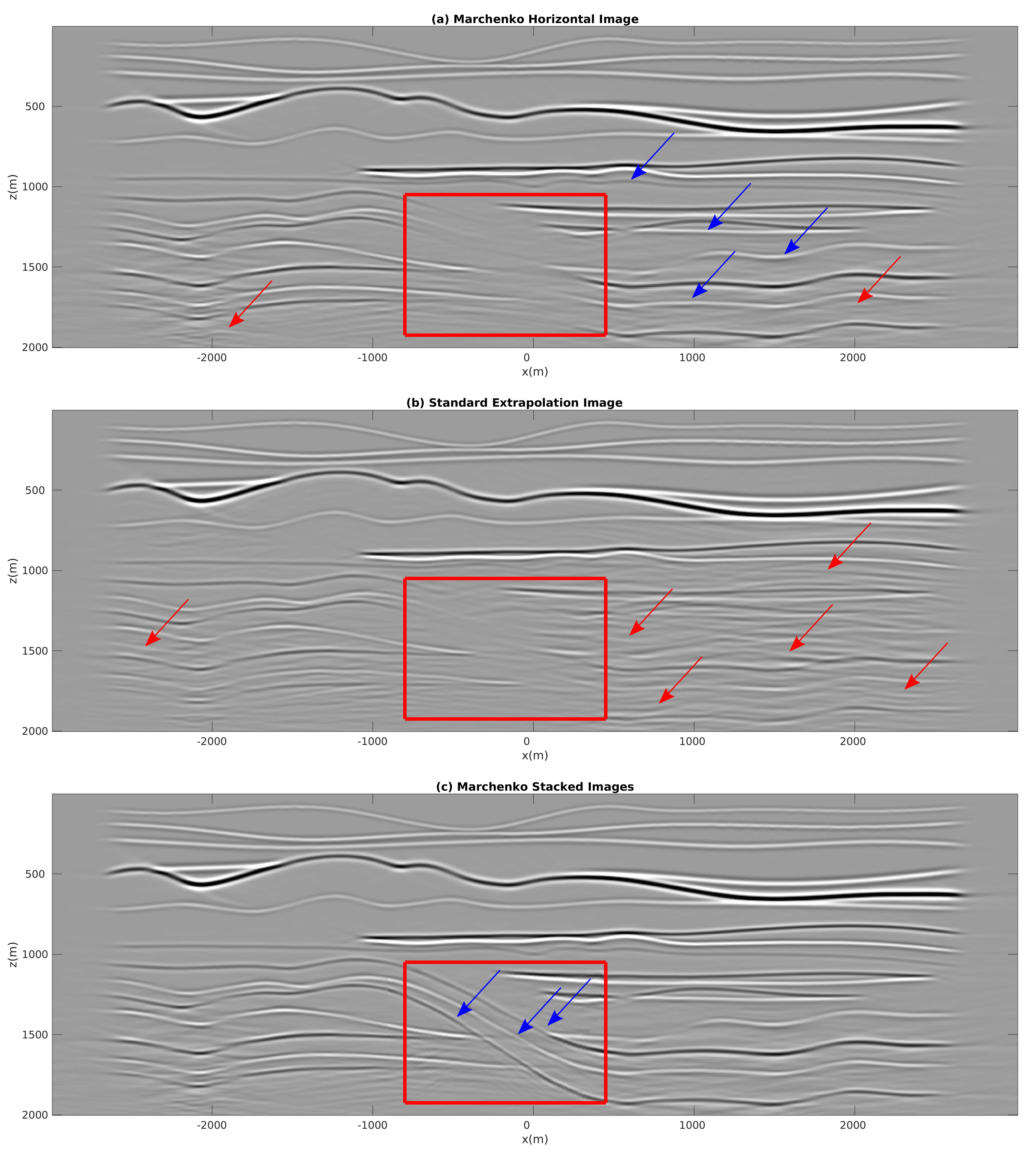} 
\caption{(a) Migration result using the imaging
condition of equation (\ref{migration_imaging}) and Marchenko redatumed virtual-plane wavefields. The red arrows point at low amplitude artefacts, 
whereas the blue arrows point at resolved structures not visible in the standard migration image.  The red box encircles an area where dipping interfaces are not imaged.
(b) Migration result using standard one-way extrapolation of virtual-plane wavefields. Red arrows point at multiple-related artefacts.
(c) Migration result using plane wave Marchenko wavefields associated with the tilted planes in Figure \ref{2D_model}(c). Blue arrows indicate dipping interfaces now properly imaged.}
\label{Imaging_2}
\end{figure}

The migration associated with the imaging condition in equation (\ref{migration_imaging}) is shown in Figure \ref{Imaging_2}(a). 
While only minor multiple-related artefacts contaminate the migration result (red arrows in Figure \ref{Imaging_2}(a)), some dipping interfaces are not imaged (red box in Figure \ref{Imaging_2}(a)).
However, most structures are properly identified. As for the first experiment, multiple-related artefacts contaminate the image if we migrate 
standard one-way extrapolated wavefields (red arrows in Figure \ref{Imaging_2}(b)). Blue arrows in Figure \ref{Imaging_2}(a) point at structures clearly visible in the Marchenko image that 
are partially or totally overshadowed by coherent noise in the standard one-way extrapolation result.
Whereas the occurrence of only minor false positives (i.e., multiple-related artefacts) testifies the potential of the method, the presence of false negatives (i.e., 
the unability to image dipping interfaces) also indicates  its limitations.
The dipping structures in the red box are not imaged due to the poor illumination provided by horizontal areal-sources.
Despite this limitation, the method provides an unexpensive multiple-related free image which could be used to guide more expensive target imaging methods to areas of interests otherwise
overshadowed by coherent noise in the standard one-way extrapolation results (such as that of Figure \ref{Imaging_2}(b)).

Moreover, we can improve the imaging results at a small cost by simply employing dipping boundaries in the subsurface and retrieving, 
via solution of an appropriate Marchenko equation, in- and out-propagating 
areal-sources-responses associated with tilted planes (the derivation of these new equations is obtained by simply replacing $\Lambda_f$  in equation (\ref{ComplexMarchenko}) with a tilted boundary,  
and replacing downgoing and upgoing fields by outward and inward propagating fields). 
The retrieved wavefields provide better illumination of dipping interfaces and therefore improve the overall quality of the migration.
To emphasize the benefit of this strategy, we target the area in the red box and employ 4 sets of dipping boundaries (white and black lines in Figure \ref{2D_model}(c)). 
For the geometry considered here, 600 additional Marchenko solutions are used. The same procedure as discussed above, consisting 
of Marchenko areal-source-response estimation and migration, is then followed.
The images associated with horizontal and dipping boundary migrations are finally stacked together and the result is shown in Figure \ref{Imaging_2}(c). 
While no false positive is contaminating the final image, the previously invisible tilted interfaces are now
visible (blue arrows in Figure \ref{Imaging_2}(c)). Consider that the area of the target zone alone is about $1 km^2$, which would correspond to about 40000 virtual source locations (at a 5
meters sampling rate) if we chose to image it with 
other Marchenko schemes. Thus, while involving dipping boundaries increases the computation burden of the areal-source method presented here, 
its convenience with respect to point-source methods is still significant.\\

\begin{figure} 
\centering
\includegraphics[width=0.95\textwidth]{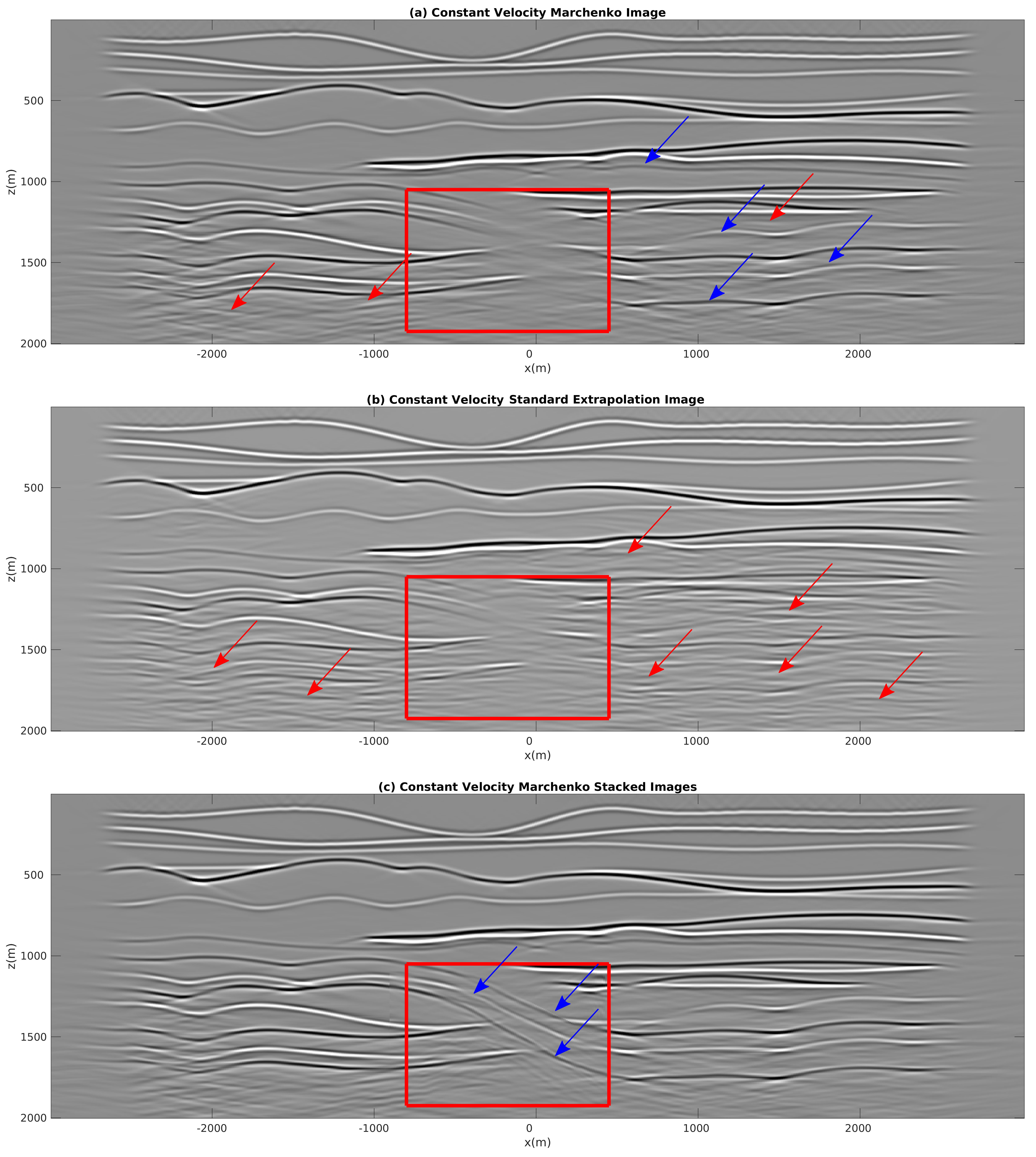} 
\caption{As for Figure (\ref{Imaging_2}), but now using a simplified, constant velocity model in the redatuming and migration steps.}
\label{Imaging_3k}
\end{figure}

In the examples discussed above we have employed smoothed versions of the actual velocity/density models in the redatuming and migration steps.
However, in practical situations it is not always possible to have access to such good estimates of the true model parameter distributions.
To assess the applicability of the method in realistic scenarios we therefore test the Marchenko areal-source scheme also assuming very poor prior knowledge of the actual velocity/density distributions. 
More specifically, we repeat the second experiment using homogeneous velocity/density models, setting $c=2000m/s$ and $\rho=2000 kg/m^3$. 
Several authors have already discussed the sensitivity of the Marchenko method to various source of errors (\cite{Thorbecke2013, Meles2015, Meles2016,Ridder2016}), and here 
its robustness is further tested.
Figure \ref{Imaging_3k}(a) shows the migration output corresponding to Marchenko horizontal areal-source-responses. While the image is distorted due to the simplified
velocity model employed in the migration step, only minor multiple-related artefacts contaminate the image (red arrows in Figure \ref{Imaging_3k}(a)). This demonstrates the good performance
of the plane-wave Marchenko method in suppressing standard wavefield extrapolation artefacts even if the employed velocity model is strongly simplified.
On the other hand, strong multiple-related artefacts are superimposed to actual interfaces when standard one-way extrapolated wavefields are migrated (red arrows Figure \ref{Imaging_3k}(b)).
Blue arrows in Figure \ref{Imaging_3k}(a) point at structures clearly visible in the Marchenko image that are partially or totally overshadowed by coherent noise in the one-way extrapolation result.
As for the previous case, if we include tilted boundaries in the redatuming and migration process, the dipping interfaces are imaged (Figure \ref{Imaging_3k}(c).

\section{Conclusions}

We have demonstrated that Marchenko methods can be successfully applied beyond conventional space-time focusing. We have discussed how a modified 
focusing condition relates areal-source-responses associated with horizontal or dipping planes to standard reflection data. A separation operator  based on specifically
designed direct focusing functions can then be applied to convolution/cross-correlation representation theorems to retrieve areal-source-responses 
at the surface through standard Marchenko algorithms. The retrieved wavefields can be used to produce images, free of multiple-related artefacts, at a fraction of the cost
of standard Marchenko approaches, thus potentially guiding expensive target imaging and being applicable also for 3D data-sets.
While more complex problems could deteriorate the performances of the proposed method, the results discussed above demonstrate its applicability to a large class of problems.
Analysis and assessment of the resolution properties of the proposed method with respect to standard Marchenko imaging and its extension to elasticity are topics of ongoing research.

\section{Acknowledgments}
We thank Joost van der Neut, Lele Zhang, Christian Reinicke, Evert Slob, Joeri Brackenhoff and Myrna Staring (Delft University of Technology)
for their collaboration and for fruitful discussions which inspired this paper. This work is partly funded by the European Research Council 
(ERC) under the European Union's Horizon 2020 research and innovation programme (grant agreement No: 742703). We also thank Nobuaki Fuji and Katrin L{\"o}er
whose comments have improved the quality of this manuscript.

\bibliographystyle{apalike}
\bibliography{paper}

\end{document}